\documentclass[final,5p,times,twocolumn]{elsarticle}
\biboptions{sort&compress}

\usepackage{lipsum}
\usepackage{amsmath, amssymb, graphicx, physics, bm}

\usepackage{mathrsfs} 
\usepackage{lineno}

\journal{Physics Letters B}

\begin{document}

\begin{frontmatter}



\title{Krylov Complexity and Entropy of Composite Quantum Systems}


\author[1,2]{Jeff Murugan}
\author[1]{Victor Van Aswegen}
\author[1,2]{Hendrik J R Van Zyl}
\affiliation[1]{organization={The Laboratory for Quantum Gravity \& Strings, Department of Mathematics and Applied Mathematics, University of Cape Town, Cape Town, South Africa}}
\affiliation[2]{organization={National Institute of Theoretical and Computational Sciences (NITheCS),
Stellenbosch, 7604, South Africa}}

\begin{abstract}
We investigate the behaviour of Krylov complexity as the number of degrees of freedom of a quantum system increases. Specifically, we show that for tensor products of commuting subsystems, the leading excess complexity is controlled by a weighted variance of the subsystem Lanczos coefficients. This provides a quantitative measure of dynamical heterogeneity and a precise characterization of synchronous evolution. We also show that Krylov entropy is able to disambiguate the information associated with subsystems composition from that generated by dynamical mismatch, and that the same Lanczos variance controls the leading mismatch contribution to both complexity and the entropy. Our results establish excess complexity and Krylov entropy as complementary probes of the emergence of collective dynamics in large quantum systems.
\end{abstract}



\begin{keyword}
Krylov Complexity \sep Krylov Entropy \sep Tensor Products \sep Excess Complexity 



\end{keyword}

\end{frontmatter}




\section{Introduction}
\label{introduction}
\noindent
Spread complexity (and its equivalent for quantum operators, Krylov complexity) \cite{Parker:2018yvk,Balasubramanian:2022, Nandy:2024evd, Rabinovici:2025otw, Baiguera:2025dkc} has become a prominent tool in understanding the dynamic properties of quantum systems.  In particular, these ideas (and their variations e.g. \cite{Craps:2024suj, caputa2025block, Caputa:2025ozd}) have appeared in the study of quantum chaos \cite{ Avdoshkin:2022xuw,  Balasubramanian:2022tpr, Hashimoto:2023swv, Erdmenger:2023wjg, Alishahiha:2024vbf, Baggioli:2024wbz}, thermalisation \cite{Alishahiha:2024rwm}, phase transitions and geometric descriptions of quantum evolution \cite{Caputa:2021sib, Chattopadhyay:2023fob}.  A key feature of this concept is the formulation of the dynamics in terms of the Krylov chain, a semi-infinite chain that captures the full evolution of a given seed (or operator) under the system Hamiltonian.  For hermitian Hamiltonians, the dynamics on the Krylov chain is that of a nearest-neighbour spin chain, with hopping amplitudes given by the Lanczos coefficients \cite{Lanczos1950AnIM, viswanath2008recursion}.  The Lanczos algorithm provides an iterative scheme to produce the Krylov basis vectors or (equivalently) the Lanczos coefficients.  The average distance traveled by the reference state along the Krylov chain defines its complexity.  \\ \\
The iterative nature of the algorithm gives rise to some unique mathematical features of spread complexity.  For example, it does not provide a measure of distance on the Krylov subspace \cite{Aguilar-Gutierrez:2023nyk} and, in contrast to alternative notions of complexity such as Nielsen complexity \cite{Nielsen:2012yss}, is superadditive under tensor products \cite{Murugan:2026wss}.  If, and only if, the subsystems evolve in synchrony do their complexities sum to the complexity of the combined system.  The details of the Krylov basis in tensor products are also revealing.  Even if a separable state is chosen as reference state, the Krylov basis vectors (which provide the basis that minimises the spread complexity at early times, see \cite{Balasubramanian:2022}) are in general not separable.  The condition for synchronous evolution is thus not related to separability, and the entangled nature of the Krylov basis is unavoidable.  At present, the condition for synchronous evolution is best understood in terms of the relation between the Krylov decomposition of the full system Hamiltonian and the subsystem Hamiltonians.  One of the aims of this letter is to better understand this notion by focusing on the Lanczos spectrum.  
\\ \\
These observations also raise another natural question.  As has been the belief (see \cite{Landsman:2005de} for a beautiful discussion of this) the large $N$ limit of a system with $N$ degrees of freedom should give rise to a quantum-to-chaos transition.  The details of such transitions can be subtle (see \cite{Layton:2023wdo, baldo2026classical} for some recent works  on the subject) but the intuition is rather clear;  whenever an average quantity is considered, the individual quantum fluctuations should become less relevant.  One would thus expect that, in the limit of many degrees of freedom, the dynamics should be effectively described by collective excitations and that corrections should decrease as the number of degrees of freedom increase.  A second aim of this letter is to understand how the increase in system size affects the excess complexity and in what sense (if at all) the emergence of collective behavior can be probed by it.  

\section{Excess Complexity}
\noindent
The Krylov basis is determined by a choice of Hamiltonian, $H$, and a reference state $|K_0\rangle$.  The Krylov basis $|K_n\rangle$ is then determined recursively as the linear combination of states obtained by up to $n$ applications of the Hamiltonian on the reference state orthogonal to all previous Krylov basis vectors.  Explicitly this means
$$ |K_n\rangle = N\left( H^n |K_0\rangle + \sum_{m=0}^{n-1}  c_{m}^n H^{n-1} |K_0\rangle   \right)   $$
or 
$$ |K_{n+1}\rangle = N\left( H |K_{n}\rangle + \alpha_n |K_n\rangle + \beta_{n} | K_{n-1} \rangle +  \sum_{m=0}^{n-2}  \tilde{c}_{m}^n |K_m\rangle  \right).   $$
In the case of an hermitian Hamiltonian the above simplifies to the Lanczos algorithm \cite{Lanczos1950AnIM, viswanath2008recursion}
$$ H|K_n\rangle = b_{n+1} |K_{n+1}\rangle + a_n |K_n\rangle + b_{n}|K_{n-1}\rangle. $$
In terms of the Krylov basis, a generic hermitian Hamiltonian can be thus be decomposed as 
\begin{equation}
H = L_{-} + L_0 + L_+    \label{KrylovDecomp}
\end{equation}
where 
\begin{eqnarray}
L_{+} & = & \sum_{n} b_{n+1} |K_{n+1}\rangle \langle K_n|   \nonumber \\
L_0 & = & \sum_{n} a_n |K_n\rangle \langle K_n|,    \nonumber
\end{eqnarray}
and $L_{-} = L_{+}^\dag$.  The coefficients $a_n$ and $b_n$ are the Lanczos coefficients and the Krylov basis thus casts the dynamics of an arbitrary hermitian Hamiltonian to a nearest-neighbour spin chain, the Krylov chain.  The Krylov complexity operator is then defined (see \cite{Parker:2018yvk, Caputa:2021sib, Balasubramanian:2022}) as 
$$ \hat{K} = \sum_{n} n |K_n\rangle \langle K_n|    $$
and complexity of any state in the Krylov subspace, $|\psi\rangle$ is given by $\langle \psi| \hat{K} |\psi\rangle$.  In particular, the spread complexity of the seed state $|K_0\rangle$ is defined as 
\begin{equation}
C(t) = \langle K_0| e^{i t H} \hat{K} e^{- i t H} |K_0\rangle.
\end{equation}
The spread complexity characterises the spread of the initial state through Krylov space.  A particularly insightful way to characterise this is as evolution on the semi-infinite Krylov chain \cite{Erdmenger:2023wjg} with spread complexity measuring the mean distance along the chain. \\ \\
A natural question that emerges from this perspective is what happens when the quantum system under consideration is a tensor product of commuting subsystems.  In a recent work \cite{Murugan:2026wss}, it was demonstrated that if $\hat{K}_j$ is the complexity operator sourced by a Hamiltonian $H_j$ and reference state $|K_{j,0}\rangle$ then the operator 
$$\hat{\Delta} = \hat{K}_{123\cdots n} - \oplus_{j=1}^n \hat{K}_j $$
where $\hat{K}_{123\cdots n}$ is the complexity operator for the combined system (sourced by the Hamiltonian $\oplus H_j$ and reference state $|K_{0,1}\rangle \otimes |K_{0,2}\rangle \otimes \cdots |K_{0,n}\rangle$) is semi-positive definite.  The operator $\hat{\Delta
}$ measures the \textit{excess complexity} of the tensor product as compared to the subsystems.  In particular, the semi-positive definiteness implies that the excess complexity of the time-evolved reference state
$$ \Delta(t) = C_{123\cdots n}(t) - \sum_{j=1}^n C_{j}(t)$$
is a non-negative quantity for all times.  It was furthermore shown that the excess complexity is zero if and only if the subsystems evolve \textit{synchronously}.  By this it is meant that the decomposition of the combined system (\ref{KrylovDecomp})
is matched perfectly with the decomposition of the subsystem as\footnote{Note that this statement implies that $L_{-}$ and $L_0$ is matched similarly.} 
$$ L_{+} = \oplus_{j=1}^n L_{j,+}. $$
In this letter one of our aims will be to unpack the concept of synchronous evolution in more detail by focusing on how it is realised on the level of the Lanczos coefficients.  Furthermore, we will investigate how the excess complexity is affected if a large number of subsystems is considered.  

\section{Early time expansion}
\label{earlyTimeAppendix}
\noindent
It is well-known that the early-time expansion of complexity gives an expression in terms of the Lanczos coefficients (see e.g. \cite{Aguilar-Gutierrez:2023nyk})
$$ C(t) = b_1^2 t^2 - \frac{1}{12}b_1^2\left( (a_0 - a_1)^2 + 4 b_1^2 - 2 b_2^2   \right)t^4 + \cdots.$$
In turn, the Lanczos coefficients can be expressed in terms of the moments of the Hamiltonian (see e.g. \cite{Parker:2018yvk}).  Note that the dynamics do not depend on the value of $a_0 = \langle K_0 | H | K_0\rangle$ directly, but only on the differences $a_j - a_0$.  Put differently, we can always shift the Hamiltonians by a constant so that $a_{0,i} = 0$.  We will use this freedom throughout to obtain simpler expressions.  Written in terms of the moments, the early-time complexity expands (after implementing this shift) as
{\small
\begin{eqnarray}
C(t) & = & \langle K_0 | H^2 | K_0\rangle t^2 \nonumber \\
&-& \frac{1}{6}\left( 3\langle K_0 | H^2 | K_0\rangle^2 - \langle K_0 | H^4 | K_0\rangle   +  \frac{3}{2} \frac{\langle K_0 | H^3 | K_0\rangle^2}{\langle K_0 | H^2 | K_0\rangle}   \right) t^4 + O(t^6)   \nonumber
\end{eqnarray}
}
\noindent
Let's consider the case where the Hamiltonian is a Kronecker sum of subsystem Hamiltonians.  Each of these Hamiltonians can be decomposed via the Lanczos algorithm into the form 
$$H_i = L_{+,i} + L_{0,i} + L_{-,i} - E_{0,i}$$
where $E_{0,i} = \langle K_0| L_{0,i} |K_0\rangle$.  We perform this shift so that 
\begin{equation}
a_{0,j} = \langle K_0| H_j |K_0\rangle = 0,    \label{a0isZero}
\end{equation}
which will ensure more simplified expression, since 
$$ L_{-,j} |K_0\rangle =  (L_{0,j} - E_{0,j}) |K_0\rangle = 0.$$   The system Hamiltonian is given by 
$$H = \oplus_i H_i.$$

\noindent
A systematic way to calculate the excess complexity is to note that
\begin{eqnarray}
& & \langle K_0| (H_1\otimes I + I \otimes H_2)^n | K_0 \rangle     \nonumber \\
& = & \sum_{k =0}^n \frac{n!}{(n-k)! k!} \langle K_0| (H_1^k \otimes H_2^{n-k} | K_0\rangle  
\end{eqnarray}
for commuting operators.  Using this and the simplifications in (\ref{a0isZero})
we find, for example,
\begin{eqnarray}
\langle K_0| H^2 |K_0\rangle & = & \sum_i \langle K_0| H_i^2 |K_0 \rangle    \nonumber \\
\langle K_0| H^3 |K_0\rangle & = & \sum_i \langle K_0| H_i^3 |K_0 \rangle    \nonumber \\
\langle K_0| H^4 |K_0\rangle & = & 3 \langle K_0| H^2 |K_0\rangle^2  \nonumber \\
& + & \sum_i \left( \langle K_0| H_i^4 |K_0 \rangle - 3  \langle K_0| H_i^2 |K_0 \rangle^2    \right).  \nonumber 
\end{eqnarray}
When inserting these simplifications into the expression for spread complexity we find, up to fourth order in $t$, that 
{\footnotesize
\begin{eqnarray}
& & C(t) - \sum_i C_i(t) \nonumber \\
&=& -\frac{1}{6}\left(\frac{3}{2} \frac{\langle K_0 | H^3 | K_0\rangle^2}{\langle K_0 | H^2 | K_0\rangle} - \frac{3}{2} \sum_i \frac{\langle K_0 | H_i^3 | K_0\rangle^2}{\langle K_0 | H_i^2 | K_0\rangle} \right)t^4 + O(t^6)    \nonumber \\
& = & \frac{1}{4} \frac{\sum_{ij} \left(    \frac{\langle K_0 | H_i^3 | K_0\rangle^2}{\langle K_0 | H_i^2 | K_0\rangle}\langle K_0 | H_j^2 | K_0\rangle   -  \langle K_0 | H_i^3 | K_0\rangle \langle K_0 | H_j^3 | K_0\rangle       \right)t^4}{\sum_k \langle K_0 | H_k^2 | K_0\rangle }  + O(t^6)  \nonumber \\
& = & \frac{1}{8} \frac{\sum_{ij} \left(   \langle K_0 | H_i^3 | K_0\rangle  \sqrt{\frac{\langle K_0 | H_j^2 | K_0\rangle}{\langle K_0 | H_i^2 | K_0\rangle}}   - \langle K_0 | H_j^3 | K_0\rangle  \sqrt{\frac{\langle K_0 | H_i^2 | K_0\rangle}{\langle K_0 | H_j^2 | K_0\rangle}}  \right)^2 t^4}{\sum_k \langle K_0 | H_k^2 | K_0\rangle }  + O(t^6) \nonumber 
\end{eqnarray}
}
\noindent
This quantity is always positive, in harmony with the general result.  Written in terms of the Lanczos coefficients, this reads
\begin{eqnarray}
    C(t) - \sum_i C_i(t) = \sum_{ij } \frac{b_{1,i}^2 b_{1,j}^2 }{8 \sum_k b_{1,k}^2}\left( a_{1,i} -  a_{1,j}    \right)^2 t^4 + O(t^6)\,.
    \label{eq:DeltaC}
\end{eqnarray}
This expression can be made even more transparent if we define
\begin{eqnarray}
    w_i=\frac{b_{1,i}^2}{\sum_jb_{1,j}^2}\,,    \label{Wweights}
\end{eqnarray}
in terms of which,
\begin{eqnarray}
    \sum_{ij}w_iw_j(a_i-a_j)^2
=
2\left[
\sum_iw_i a_i^2-\left(\sum_iw_i a_i\right)^2
\right]\,,
\end{eqnarray}
and the excess complexity,
\begin{eqnarray}
    \Delta C(t) = \frac14 \left(\sum_i b_{1,i}^2\right)
\mathrm{Var}_{w}(a_{1,i})\,t^4
+O(t^6)\,.
    \label{eq:DeltaC(var)}
\end{eqnarray}
In words, the coefficient of the leading excess complexity is just a weighted variance of the first diagonal Lanczos coefficients across the subsystems.\footnote{Intriguingly, symmetry resolved Krylov complexity and -spread complexity \cite{Caputa:2025mii, Caputa:2025ozd} also have early time expansions involving weighted sums of Lanczos coefficients.  The origin of the weighting is related to the makeup of the reference state and not to the Lanczos coefficients, however.} Another way of saying this is that if all the subsystems have $a_{1,i}=a_1$,
then $\mathrm{Var}_w(a_1)=0$ and 
at this order, their Krylov dynamics are synchronized; after the first Krylov step they all see the same diagonal Lanczos coefficient. Consequently, there is no $t^4$ excess. If however the $a_{1,i}$’s differ, the subsystems begin to distinguish themselves dynamically at this order, and $\mathrm{Var}_w(a_1)>0$. It is then precisely this mismatch that generates a positive excess complexity.  \\ \\
In the case\footnote{Note that this occurs when the reference state gives rise to $\langle (H_i - \langle H_i\rangle)^3 \rangle = \langle (H_i - \langle H_i\rangle)^5 \rangle = 0 $ for each subsystem.} where $a_1 =0$ and $a_2 = 0$ the leading order contribution is given by 
\begin{eqnarray}
    \Delta C(t) = \frac{1}{18} \left(\sum_i b_{1,i}^2\right)
\mathrm{Var}_{w}(2 b_{1,i}^2 - b_{2,i}^2)\,t^6
+O(t^8)\,,
  \label{eq:DeltaCt6}
\end{eqnarray}
which is again a sum of strictly positive terms, in harmony with the general result.  \\ \\
Finally, in the case that $a_{1,i}$ are all equal to $a_1$ but non-zero then the $O(t^4)$ excess is again zero.  The $O(t^6)$ excess now gains contributions from terms that can be characterised by two weightings.  These are (\ref{Wweights}) and 
\begin{equation}
v_i = \frac{ b_{1,i}^2 b_{2,i}^2      }{   \sum_j  b_{1,j}^2 b_{2,j}^2    } 
\end{equation}
in terms of which
{\footnotesize
\begin{eqnarray}
  & &   \Delta C(t)   \nonumber \\
  & = & \frac{1}{18} \left(\sum_i b_{1,i}^2\right)
\mathrm{Var}_{w}(2 b_{1,i}^2 - b_{2,i}^2)\,t^6  \nonumber  \\
& & +    \frac{1}{18} \left(\sum_i b_{1,i}^2 b_{2,i}^2  \right) \left(  \sum_{i \neq j} b_{1,i}^2 b_{1,j}^2   \right) \frac{  \left( \mathrm{Avg}_{v}(2 a_{1} - a_{2,i})  \right)^2 }{      \left(\sum_i b_{1,i}^2 b_{2,i}^2  \right) + 2 \sum_{i \neq j} b_{1,i}^2 b_{1,j}^2   }\,t^6       \nonumber  \\
& &  +    \frac{1}{36} \left(\sum_i b_{1,i}^2 b_{2,i}^2  \right) \mathrm{Var}_{v}(2 a_{1} - a_{2,i})\,t^6  
+O(t^8)\,. \label{deltaSpecial}
\end{eqnarray}
}
One may have guessed that one of the conditions for synchronicity is that the Lanczos coefficients for the subsystems must be equal.  Eq. (\ref{deltaSpecial})  shows that this is not case once diagonal entries are included.  If the $a_{2,i}$ are not zero then we require a stricter condition, namely that $a_{2,i} = 2 a_1$.  Synchronicity (even at $O(t^6))$ thus not only requires that the Lanczos coefficient be equal but that a particular ratio of the coefficients be satisfied.  Going to higher orders confirms this pattern.  Only specific algebraic structures (such as that of the Heisenberg-Weyl algebra) give rise to vanishing excess complexity - see \ref{HeisenbergAlgAppendix}.  \\ \\
Furthermore, we note that the excess complexity (both at $O(t^4)$ and $O(t^6)$) scales with the number of elements in the tensor product.   This is evident, for example, from the $\sum_i b_{1,i}^2$ factors.  The ratio between the excess complexity and the total system complexity at early times is given by 
$$ \frac{\Delta C(t)}{C(t)} = \frac{1}{4} \mathrm{Var}_{w}(a_{1,i})\,t^2  + O(t^4) $$
which is an $O(1)$ number.  We thus conclude that the excess complexity remains significant even as the number of tensor products is increased.  The only exception to this is in the early-time regime when the total system complexity still grows quadratically in time.  To see this, define 
$$ t = \frac{\tau}{ \sqrt{ \sum_i b_{1,i}^2  }  } $$
in terms of which 
$$ C(\tau) = \tau^2 + o(\tau^4)   $$
In terms of $\tau$ the leading order contribution for excess complexity scales inversely proportional to the number of subsystems.  Within this time-scale the complexity of the total system is thus close to synchronous evolution in a perturbative sense.  It is important to note, however, that this time-scale that defines the early-time growth of complexity itself is inversely proportional to the number of subsystems.

\section{Krylov entropy and subsystem composition}
\label{sec:krylov_entropy}
\noindent
The excess complexity analysed in the previous section measures the
departure of the combined Krylov dynamics from the sum of the
individual subsystem complexities. It is useful to complement this
with an information-theoretic characterization of the corresponding
Krylov distributions. Specifically, for a state $|\psi(t)\rangle=\sum_n \phi_n(t)|K_n\rangle$, we can define the Krylov entropy (see \cite{Barbon:2019wsy})
\begin{equation}
    S_K(t)
    =
    -\sum_n p_n(t)\log p_n(t)\,,
    \qquad
    p_n(t)=|\phi_n(t)|^2.
    \label{eq:Krylov_entropy}
\end{equation}
Consider, as in the previous section, a collection of commuting
subsystems. The product Krylov basis is labelled by $\bm n=(n_1,\ldots,n_N)$ with probability distribution
\begin{equation}
    P_{\bm n}(t)
    =
    \prod_i p^{(i)}_{n_i}(t).
\end{equation}
Its entropy is therefore exactly additive,
\begin{equation}
    S_K^{\rm prod}(t)
    =
    -\sum_{\bm n}P_{\bm n}\log P_{\bm n}
    =
    \sum_i S_K^{(i)}(t).
    \label{eq:product_entropy}
\end{equation}
There is a useful decomposition of this entropy which makes its
relation to the combined Krylov chain transparent. Define the radial
Krylov coordinate $r=\sum_i n_i$,
and let $P_r$ denote the probability of occupying the corresponding shell.  The chain rule for Shannon entropy (\cite{Shannon:1948dpw}) gives
\begin{equation}
    S_K^{\rm prod}
    =
    S_\parallel+S_\perp,
    \label{eq:entropy_decomposition}
\end{equation}
where
\begin{equation}
    S_\parallel
    =
    -\sum_r P_r\log P_r
\end{equation}
is the entropy associated with spreading between radial shells, while
\begin{equation}
    S_\perp
    =
    \sum_r P_r
    \left[
        -\sum_{\bm n:\,\sum_i n_i=r}
        P(\bm n|r)\log P(\bm n|r)
    \right]
    \label{eq:Sperp_def}
\end{equation}
measures the information contained in the distribution within each
shell.  Thus $S_\perp$ records information about which subsystem
carries a given Krylov excitation that is absent from a purely radial
description. \\

\noindent
To make this explicit at early times, we need to use shift-invariant Lanczos data. Toward this end, we define
\begin{equation}
    \alpha_i
    \equiv
    a_{1,i}-a_{0,i},
    \qquad
    x_i=b_{1,i}^2,
    \qquad
    y_i=b_{2,i}^2.
    \label{eq:shift_invariant_data}
\end{equation}
Equivalently, $\alpha_i$ is the first diagonal Lanczos coefficient
after shifting the subsystem Hamiltonian so that $a_{0,i}=0$.
The early-time probabilities are
\begin{align}
    p^{(i)}_0(t)
    &=
    1-x_i t^2
    +
    x_i
    \left(
        \frac{\alpha_i^2}{12}
        +\frac{x_i}{3}
        +\frac{y_i}{12}
    \right)t^4
    +O(t^6),
    \label{eq:p0early}
    \\
    p^{(i)}_1(t)
    &=
    x_i t^2
    -
    x_i
    \left(
       \frac{\alpha_i^2}{12}
        +\frac{x_i+y_i}{3}
    \right)t^4
    +O(t^6),
    \label{eq:p1early}
    \\
    p^{(i)}_2(t)
    &=
    \frac{x_i y_i}{4}t^4+O(t^6).
    \label{eq:p2early}
\end{align}
Introducing
\begin{equation}
    B=\sum_i x_i,
    \qquad
    w_i=\frac{x_i}{B},
    \label{eq:entropy_weights}
\end{equation}
the first shell has total probability
$ P_1(t)=Bt^2+O(t^4)$, while, conditioned on being in that shell,
$P(1_i|r=1) = w_i+O(t^2).$ It follows immediately that
\begin{equation}
    S_\perp(t)
    =
    B H(w)t^2+O(t^4)\,,    
    \label{eq:Sperp_early}
\end{equation}
where $H(w)=-\sum_iw_i\log w_i$.
This leading contribution is present even when all subsystems evolve
synchronously.  It is kinematic rather than a measure of
dynamical mismatch in the sense that it simply counts the information associated with the different subsystem directions within the first Krylov shell. To compare the entropy of the combined Lanczos chain with that of the product Krylov space we define
\begin{equation}
    \Delta S_K(t)
    \equiv
    S_K^{\rm comb}(t)-S_K^{\rm prod}(t).
    \label{eq:DeltaSK}
\end{equation}
At leading order the combined chain has a single first Krylov
direction with $ b_{1,\rm comb}^2=B$.
Consequently,
\begin{equation}
    \Delta S_K(t)
    =
    -BH(w)t^2+O(t^4\log t)
    =
    -S_\perp(t)+O(t^4\log t).
    \label{eq:DeltaSKleading}
\end{equation}
The negative sign here comes from the fact that the tensor-product
description distinguishes the different microscopic directions in
which the state may leave the reference state, whereas the combined
Lanczos construction compresses these directions into a single
collective Krylov direction. The first contribution that is sensitive to dynamical mismatch enters
at the next order.  The first diagonal coefficient of the combined
chain is
\begin{equation}
    \alpha_{\rm comb}
    =
    \bar\alpha
    \equiv
    \sum_iw_i\alpha_i,
    \label{eq:alpha_comb}
\end{equation}
while its second hopping coefficient is
\begin{equation}
    b_{2,\rm comb}^2
    =
    \mathrm{Var}_w(\alpha)
    +
    \sum_iw_i y_i
    +
    2
    \left(
        B-\sum_iw_i x_i
    \right),
    \label{eq:b2combined}
\end{equation}
where
\begin{equation}
    \mathrm{Var}_w(\alpha)
    =
    \sum_iw_i\alpha_i^2
    -
    \left(\sum_iw_i\alpha_i\right)^2.
\end{equation}
In other words, precisely the same weighted variance that controls the leading excess complexity appears as an additional contribution to the second hopping coefficient of the combined Krylov chain. Indeed, expanding the entropy of a general Krylov chain gives
{\footnotesize
\begin{align}
    S_K(t)
    ={}&
    x t^2\left[1-\log(xt^2)\right]
    \nonumber\\
    &+
    t^4
    \left[
        x\left(
            \frac{\alpha^2}{12}
            +\frac{x+y}{3}
        \right)\log(xt^2)
        -\frac{x^2}{2}
        +\frac{xy}{4}
        \left(
            1-\log\frac{xyt^4}{4}
        \right)
    \right]
    +O(t^6\log t). \label{eq:single_chain_entropy_expansion}
\end{align}
}
Using (\ref{eq:b2combined}), the part of
$\Delta S_K$ generated by the mismatch of the diagonal Lanczos
coefficients is
\begin{equation}
    \delta_\alpha\Delta S_K(t)
    =
    \frac{B}{2}
    \operatorname{Var}_w(\alpha)
    t^4\log\frac{1}{t}
    +O(t^4).
    \label{eq:entropy_mismatch}
\end{equation}
Here the argument of the logarithm is understood to contain an
appropriate microscopic energy scale; the coefficient of
$\log t$ is independent of this choice. The excess complexity found in the previous section may be written in the explicitly shift-invariant form
\begin{equation}
    \Delta C(t)
    =
    \frac{B}{4}
    \operatorname{Var}_w(\alpha)
    t^4+O(t^6).
    \label{eq:DeltaC_shift_invariant}
\end{equation}
Consequently,
\begin{equation}
    \delta_\alpha\Delta S_K(t)
    =
    2\Delta C(t)\log\frac{1}{t}
    +O(t^4).
  \label{eq:entropy_complexity_relation}
\end{equation}
Equations (\ref{eq:DeltaSKleading}) and
(\ref{eq:entropy_complexity_relation}) separate two distinct effects.
The leading entropy deficit is combinatorial and is already present
for synchronous subsystems.  Dynamical heterogeneity first enters at
order $t^4\log t$ and is governed by the same weighted variance of the
shift-invariant Lanczos coefficients that controls the leading excess
complexity.  Krylov entropy and excess complexity therefore probe
complementary aspects of the emergence of collective Krylov dynamics; the former quantifies the loss of subsystem-resolved information, while the latter measures the dynamical cost of asynchronous evolution.

\section{The Kitaev Chain}
\label{Kitaev}
\noindent
We employ the Kitaev chain \cite{Kitaev:2001cha} to demonstrate the results detailed above. The model consists of a scalable 1D system of Majorana fermions that together exhibit striking global features such as quantum phase transitions \cite{Caputa:2023spr} – despite the relative mathematical simplicity of the model itself. Because of this, it serves as a strong mathematical probe of emergent behaviour from simple constituents. As shown in \cite{Caputa:2023spr}, the original Hamiltonian devised by Kitaev 
\begin{eqnarray}
H & = & \sum_{j=1}^{L} \left[
-\frac{J}{2} \left( c_j^\dagger c_{j+1} + c_{j+1}^\dagger c_j \right)
- \mu \left( c_j^\dagger c_j - \frac{1}{2} \right)   \right.    \nonumber \\
& & \left. + \frac{1}{2} \left( \Delta c_j^\dagger c_{j+1}^\dagger + \Delta^* c_{j+1} c_j \right)
\right],
\end{eqnarray}
(where $c_j$ and $c_j^\dagger$ are the fermionic annihilation and creation operators respectively, $J$ describes nearest-neighbour hopping, $\mu$ the chemical potential, and $\Delta$ the strength of the superconducting gap), can be Fourier-transformed under $$c_j = \frac{1}{\sqrt{L}} \sum_{n=0}^{L-1} e^{ik_nj} a_{k_n}.$$  Assuming even $L$ and anti-periodic boundary conditions\footnote{If this is not satisfied, additional terms for momenta $k = 0$ and $k = \pi$ should be added to the Hamiltonian.  } i.e. $c_1=-c_{L+1},$ the expression for the Hamiltonian becomes \begin{eqnarray}    
H& = &-\sum_{n=0}^{L/2- 1}\left[2(\mu+J\cos(k_n))J_0^{(k_n)}-i\Delta\sin(k_n)(J_+^{(k_n)}-J_-^{(k_n)})\right].   \nonumber
\\
k_m & = & \frac{2 \pi}{L}\left( m + \frac{1}{2}  \right) \label{Htransformed}
\end{eqnarray}
where $J_0^{(k_n)},$ $J_-^{(k_n)}$ and $J_+^{(k_n)}$ satisfy an $su(2)$ algebra and are defined as quadratic combinations of the $a_{k_n},$ $a_{k_n}^\dagger$ \cite{Caputa:2022yju}. A reference state is then the only remaining element required to find the Krylov complexity. In this letter we will choose a tensor product of $L/2$ (identical) lowest weight eigenstates of $J_0^{(k_n)},$ individually written as $\ket{s,-s}$ and specified by $$J_-^{(k_n)}\ket{s,-s}=0 \qquad \text{and} \qquad J_0^{(k_n)}\ket{s,-s}=-s\ket{s,-s}.$$
The combined state of the system is then labelled by $\ket{j,-j}$ and written as $$\ket{j,-j}=\otimes_{{k_n}=0}^{L/2-1}\ket{s,-s}.$$
We then set out to calculate the Lanczos coefficients of each subsystem characterised by its own Hamiltonian, corresponding to the individual term $H_i$ in the sum $(\ref{Htransformed}).$ This can be done by defining the specific two-point function $\phi_{0,n} = \langle j,-j | e^{itH_n} | j,-j \rangle,$ and implementing the corresponding method developed in \cite{Balasubramanian:2022} to find $a_{0,n},$ $a_{1,n},$ $b_{1,n},$ and $b_{2,n}.$ For each of these this requires evaluating derivatives of $\phi_{0,n}(t)$ at $t=0,$ and is explicitly calculated by using the group identity $e^{itH_{n}}=e^{\alpha_n J_+^{(k_n)}}e^{\gamma_n J_0^{(k_n)}}e^{\alpha'_n J_-^{(k_n)}}$ for constants $\alpha_n,$ $\alpha'_n$ and $\gamma_n.$ The constants are calculated within a specific choice of representation (in our case the Pauli spin matrices), shown in more detail in~\ref{Pauli}. This puts the defined two-point function into the form 
\begin{eqnarray}
   \langle j,-j|e^{itH_n}|j,-j\rangle= \exp\left\{2s\ln[C_{k_n}(t)+i[\mu+J\cos(k_n)]S_{k_n}(t)]\right\}\label{phi},
       \label{eq:kitaev_return}
\end{eqnarray} 
with 
$$
C_{k_n}\equiv\cos\!\left(t\sqrt{(\mu+J\cos(k_n))^2+\Delta^2\sin^2(k_n)}\right)$$$$S_{k_n}\equiv\frac{\sin\!\left(t\sqrt{(\mu+J\cos(k_n))^2+\Delta^2\sin^2(k_n)}\right)}{\sqrt{(\mu+J\cos(k_n))^2+\Delta^2\sin^2(k_n)}}.
$$
Taking $s=1/2$ since we are dealing with Majorana fermions, the first three nonzero Lanczos coefficients for the Hamiltonian $H_n$ are given by: 
\begin{align}
    &a_{0,n}=[\mu+J\cos(k_n)]  \nonumber  \\&a_{1,n}=[\mu+J\cos(k_n)]-2[\mu+J\cos(k_n)]   \nonumber \\&b_{1,n}=\Delta\sin(k_n).
    \label{eq:kitaev_a_a_b}
\end{align}
When implementing (\ref{eq:DeltaC(var)}) it is then seen that the excess complexity $\Delta C,$ to fourth order, is
\begin{eqnarray}
    \Delta C(t)=\frac{\Delta^2J^2L}{16}t^4+O(t^6).
    \label{eq:kitaev_leading_excess_C}
\end{eqnarray}
Furthermore, taking the case of $\mu=0$ and $J=0$ to achieve $a_{1,n}=0$ as used in (\ref{deltaSpecial}), we find that $b_{2,n}=0.$ Using this result, we can see that the second and third terms of this equation vanish by direct substitution. As a result, we recover Equation (\ref{eq:DeltaCt6}), giving \begin{eqnarray}
    \Delta C(t)=\frac{\Delta^6L}{288}t^6+O(t^8).
\end{eqnarray}

\noindent
The momentum-space decomposition of the Kitaev chain also allows the
Krylov entropy to be studied particularly explicitly.  For
$s=\frac12$, each independent $(k,-k)$ sector is a two-level Krylov
system.  Defining
\begin{equation}
    \xi_k=\mu+J\cos k,
    \qquad
    d_k=\Delta\sin k,
    \qquad
    E_k=\sqrt{\xi_k^2+d_k^2},
\end{equation}
the return amplitude in Eq.~(\ref{eq:kitaev_return}) implies that the probability of occupying the first Krylov state of the $k$th sector is
\begin{equation}
    q_k(t)
    =
    \frac{d_k^2}{E_k^2}
    \sin^2(E_k t).
    \label{eq:qk}
\end{equation}
The sector Krylov entropy is therefore simply the binary entropy
\begin{equation}
    S_{K,k}(t)
    =
    h_2(q_k)
    =
    -q_k\log q_k
    -(1-q_k)\log(1-q_k),
    \label{eq:kitaev_sector_entropy}
\end{equation}
while the entropy of the full product Krylov space is exactly
\begin{equation}
    S_K^{\rm prod}(t)
    =
    \sum_{k>0}h_2(q_k(t)).
    \label{eq:kitaev_product_entropy}
\end{equation}
In the thermodynamic limit this results in the entropy density
\begin{equation}
    \frac{S_K^{\rm prod}(t)}{L}
    \longrightarrow
    \frac{1}{2\pi}
    \int_0^\pi dk\,
    h_2
    \left[
        \frac{\Delta^2\sin^2k}{E_k^2}
        \sin^2(E_k t)
    \right].
    \label{eq:kitaev_entropy_density}
\end{equation}
At early times,
\begin{equation}
    q_k(t)
    =
    d_k^2t^2
    -
    \frac{d_k^2E_k^2}{3}t^4
    +O(t^6),
\end{equation}
so that
\begin{align}
    S_{K,k}(t)
    ={}&
    d_k^2t^2
    \left[
        1-\log(d_k^2t^2)
    \right]
    \nonumber\\
    &+
    t^4
    \left[
        \frac{d_k^2E_k^2}{3}
        \log(d_k^2t^2)
        -\frac{d_k^4}{2}
    \right]
    +O(t^6\log t).
    \label{eq:kitaev_sector_entropy_early}
\end{align}
For the anti-periodic momenta used here,
\begin{equation}
    B
    =
    \sum_{k>0}b_{1,k}^2
    =
    \Delta^2\sum_{k>0}\sin^2k
    =
    \frac{\Delta^2L}{4},
    \label{eq:kitaev_B}
\end{equation}
with corresponding weights
\begin{equation}
    w_k
    =
    \frac{b_{1,k}^2}{B}
    =
    \frac{4}{L}\sin^2k.
    \label{eq:kitaev_weights}
\end{equation}
Their entropy at large $L$ then behaves as
\begin{equation}
    H(w)
    =
    \log L-1+o(1).
    \label{eq:kitaev_weight_entropy}
\end{equation}
It follows from Eq.~(\ref{eq:Sperp_early}) that
\begin{equation}
    S_\perp(t)
    =
    \frac{\Delta^2L}{4}
    (\log L-1)t^2
    +O(t^4).
    \label{eq:kitaev_Sperp}
\end{equation}
Correspondingly, the difference between the entropy of the combined
Lanczos chain and that of the product Krylov space is
\begin{equation}
    \Delta S_K(t)
    =
    -\frac{\Delta^2L}{4}
    (\log L-1)t^2
    +O(t^4\log t).
    \label{eq:kitaev_DeltaS}
\end{equation}
This logarithmic enhancement can be understood from the fact that as the
number of momentum sectors increases, the first product-space Krylov shell contains an increasing number of distinguishable directions,
whereas the combined Lanczos construction represents their collective motion by a single radial direction. We stress that the expansion (\ref{eq:kitaev_DeltaS}) is not uniform in $L$, since its validity requires $Bt^2\ll1$, or equivalently $\Delta^2Lt^2\ll1$.
The dynamical contribution is controlled by the shift-invariant
diagonal Lanczos coefficient.  From Equations.~(\ref{eq:kitaev_a_a_b}) we find that
\begin{equation}
    \alpha_k
    =
    a_{1,k}-a_{0,k}
    =
    -2(\mu+J\cos k).
    \label{eq:kitaev_alpha}
\end{equation}
Using the weights (\ref{eq:kitaev_weights}), gives $    \langle\cos k\rangle_w=0$, and $\langle\cos^2 k\rangle_w=\frac14$,
so that
\begin{equation}
    \mathrm{Var}_w(\alpha_k)=J^2.
    \label{eq:kitaev_variance}
\end{equation}
This gives the leading excess complexity \eqref{eq:kitaev_leading_excess_C}
and, from Eq.~(\ref{eq:entropy_mismatch}), the mismatch-dependent
entropy contribution
\begin{equation}
    \delta_\alpha\Delta S_K(t)
    =
    \frac{\Delta^2J^2L}{8}
    t^4\log\frac1t
    +O(Lt^4)
    = 2\Delta C(t)\log\frac1t
    +O(Lt^4).
    \label{eq:kitaev_entropy_complexity}
\end{equation}
These results exhibit two parametrically distinct contributions to
the entropy difference.  The leading $O(L\log L\,t^2)$ contribution
is kinematic and measures the subsystem information compressed by the
combined Lanczos chain.  By contrast, the $O(Lt^4\log t)$ contribution
is dynamical and is controlled by the same weighted Lanczos variance
that generates the excess complexity.\\

\begin{figure}
    \centering
    \includegraphics[width=0.6\linewidth]{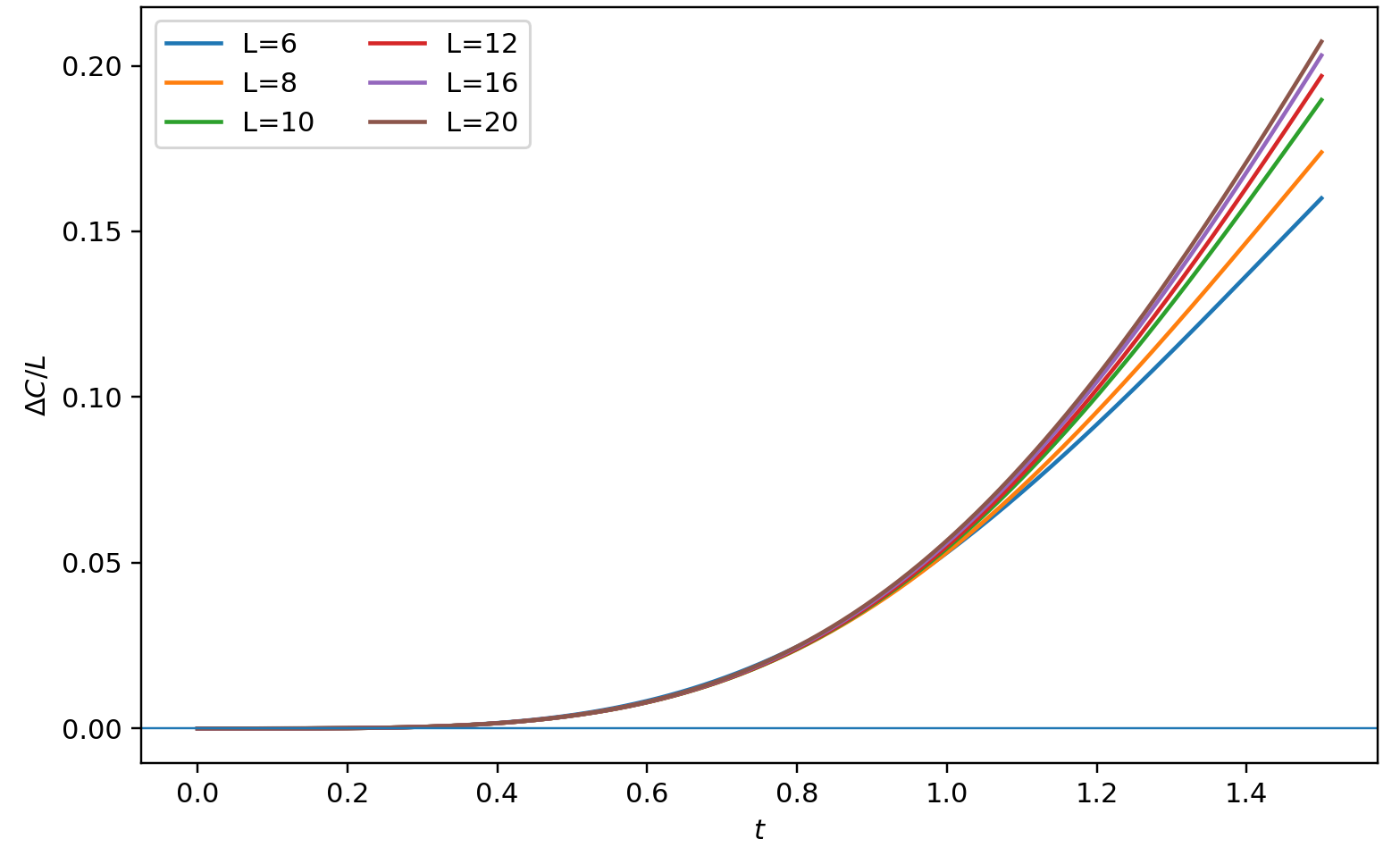}
   \caption{
Excess Krylov complexity per site for the Kitaev chain with
$\mu=J=\Delta=1$ and several system sizes.  The approximate collapse
of the curves demonstrates the extensive scaling of the excess
complexity beyond the strict early-time regime.
}
\label{fig:kitaev_complexity_numeric}
\end{figure}

\begin{figure}
    \centering
    \includegraphics[width=0.6\linewidth]{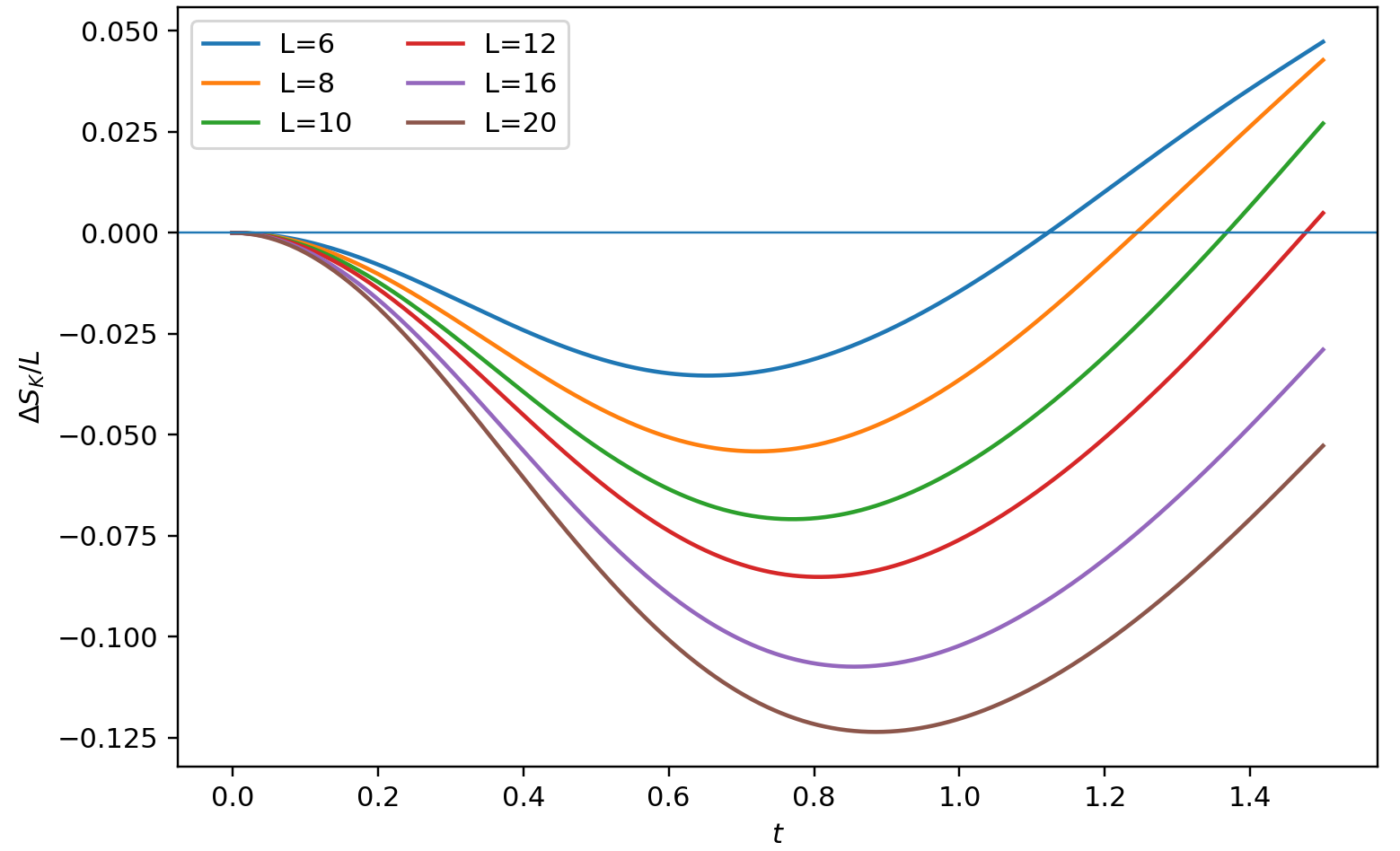}
   \caption{
Difference between the entropy of the combined Lanczos chain and the
product Krylov entropy,
$\Delta S_K=S_K^{\rm comb}-S_K^{\rm prod}$, per site for the Kitaev
chain with $\mu=J=\Delta=1$.  The negative entropy difference at early
and intermediate times reflects the compression of subsystem-resolved
Krylov information by the combined Lanczos construction.  Its
increasing magnitude with system size is consistent with the
$L\log L$ enhancement of the leading short-time contribution.
}
\label{fig:kitaev_entropy_numeric}
\end{figure}

\noindent
We can test these conclusions by constructing the Lanczos chain of the
full Kitaev Hamiltonian numerically and comparing its entropy and
complexity with those obtained from the independent momentum sectors.
Figure~\ref{fig:kitaev_complexity_numeric} shows the excess complexity
per site for several system sizes.  At early times the numerical
curves reproduce $\Delta C/(Lt^4)\rightarrow\Delta^2J^2/16$, while the approximate
collapse of $\Delta C/L$ at later times indicates that the excess
complexity remains approximately extensive beyond the regime in which
the short-time expansion is valid.
Figure~\ref{fig:kitaev_entropy_numeric} compares the combined and
product Krylov entropies.  The product entropy is systematically
larger over the early- and intermediate-time regimes, reflecting the
additional information carried by the subsystem-resolved product
basis.  Correspondingly, $\Delta S_K<0$, with the entropy deficit
becoming increasingly pronounced as the system size grows.  This
finite-size behaviour is consistent with the logarithmic enhancement
predicted by Eq.~(\ref{eq:kitaev_DeltaS}).  At later times the smaller
systems exhibit finite-size crossings; we do not attach significance
to these crossings in the thermodynamic limit.

\section{Discussion}
\noindent
In this letter we have studied the Krylov dynamics of composite quantum
systems from two complementary perspectives. Excess complexity quantifies the failure of the full-system Krylov evolution to decompose into the evolutions of its constituent subsystems, while Krylov entropy retains information about how the state is distributed among those subsystems. Together, these results provide a more detailed picture of composition in
Krylov space than either quantity provides separately.\\

\noindent
A central outcome of our analysis is that there are two distinct sources of non-additive behaviour. The first is associated purely with subsystem composition. Even perfectly synchronous subsystems possess distinguishable directions in the product Krylov space, and this information is not retained
when the dynamics is represented on the one-dimensional Krylov chain of the full Hamiltonian. The second is genuinely dynamical and arises when the constituent systems explore their respective Krylov spaces differently. The fact that the same weighted variance of the Lanczos data controls both the leading excess complexity and the first mismatch-dependent contribution
to the entropy shows that these are not unrelated effects. Rather, complexity and entropy provide dynamical and information-theoretic views of the same departure from synchronous subsystem evolution.\\

\noindent
The Kitaev chain illustrates an important consequence of this distinction. Increasing the number of momentum sectors does not make the discrepancy between the subsystem and full-system descriptions disappear. The excess
complexity remains extensive, while the leading entropy deficit acquires an
additional logarithmic enhancement with system size. The latter reflects the growing number of subsystem-resolved directions available in the product Krylov space. Consequently, at least in this example, increasing system size produces a description that is increasingly economical in terms of Krylov degrees of freedom without making the underlying subsystem structure dynamically irrelevant. This also cautions against identifying the large-system limit
with exact synchronous evolution.
These observations suggest that the relation between the product Krylov
space and the Krylov chain of the full Hamiltonian deserves further study in its own right. The shell decomposition introduced here suggests viewing the
latter as a reduced description in which information within fixed radial
shells is no longer resolved. Whether this relation can be formulated as a precise coarse-graining map, and under what conditions such a reduction becomes dynamically accurate, are natural questions raised by the present
results. In particular, it would be useful to identify a criterion for when subsystem heterogeneity becomes negligible relative to the total Krylov dynamics in a large-system limit.\\

\noindent
There are also two immediate extensions of the present work. First, the conditions for exact synchronicity remain to be classified. Our early-time results and the Heisenberg--Weyl examples of Appendix~A indicate that
vanishing excess complexity requires special relations among the Lanczos
coefficients, suggesting an underlying algebraic characterization. It would be interesting to determine the full class of Hamiltonian--reference-state
pairs for which the excess complexity vanishes identically. Second, the analysis here has relied on commuting subsystems. For genuinely interacting subsystems the product distribution no longer factorizes, so correlations
between constituents should enter both the complexity and entropy. Extending the excess-complexity construction to this setting would provide a natural test of how much of the structure uncovered here survives in interacting
many-body systems. More generally, our results suggest that composition is itself a nontrivial operation in Krylov space. Passing from subsystem-resolved dynamics to the Krylov chain of the composite system changes both the amount of information that is retained and the way in which dynamical differences between constituents are represented. Excess complexity and Krylov entropy provide complementary diagnostics of these two effects, and offer a framework for studying how the dynamics of a composite quantum system is related to that
of its constituent parts.


\section*{Acknowledgements}
JM and HJRVZ
are supported in part by the ``Quantum Technologies for Sustainable Development" grant
from the National Institute for Theoretical and Computational Sciences of South Africa
(NITHECS).

\appendix

\section{Heisenberg algebra}

\label{HeisenbergAlgAppendix}

The excess complexity is always zero if the subsystems are all elements of the Heisenberg algebra and the reference state is a Glauber coherent state.  In this appendix we will demonstrate this fact.  Consider the following system of Hamiltonians 
$$ H_j = \alpha_j a_j^\dag + \bar{\alpha}_j a_j  $$
satisfying the usual bosonic commutation relations
$$ [a_j, a_k^\dag]  = \delta_{jk} $$
First, consider the reference state for all the subsystems to be the lowest weight state 
$$ a_j |K_{0,j}\rangle  = 0  $$
It is easy to see that the Krylov basis for the subsystems is given by (\cite{Caputa:2021sib, Chattopadhyay:2023fob} )
$$ |K_{n,j}\rangle = \frac{(a^\dag_{j})^n}{\sqrt{ n! }} |K_{0,j}\rangle   $$
The complexity operator is, in fact, the number operator 
$$  \hat{K}_j = a^\dag_j a_j.   $$
The combined Hamiltonian is given by
$$H = \sum_{j} ( \alpha_j a_j^\dag    +  \bar{\alpha}_j a_j  ).      $$
Since 
\begin{eqnarray}
 \sum_j \bar{\alpha}_j a_j  |K_0\rangle & = & 0    \nonumber \\
 \left[  \sum_j \bar{\alpha}_j a_j ,  \sum_k \alpha_k a_k     \right] & = & \sum_j |\alpha_j|^2  ,   \nonumber 
\end{eqnarray}
it follows that 
$$ |K_n\rangle = \frac{ \left(  \sum_k \alpha_k a_k     \right)^n   }{ \sqrt{n!}  \sum_j |\alpha_j|^2     }  |K_0\rangle.   $$
By definition, we have that $ \langle K_n | \hat{K}_{123\cdots n}    |K_n\rangle  = n  $ but it also immediately follows that 
$$   \langle K_n| \oplus_{j=1}^n \hat{K}_j     | K_n\rangle = n$$
Taken together, this implies that the excess complexity is zero for any state in the Krylov subspace.  Importantly, this is true for any choice of constants $\alpha_j$.  \\ \\
Now consider a more general reference state, the coherent state 
$$  |z_j \rangle = N  e^{z_j a^\dag_j } |0\rangle . $$
It is well-known that these states are eigenstates of the annihilation operator $a_j$.  Put differently, the coherent states are annihilated by $ (a_j - z_j) $.  Since 
$$ \left[ a_j - z_j, a_k^\dag - \bar{z}_k    \right]  = \delta_{jk} , $$
it follows that the excess complexity for the reference state-Hamiltonian pair 
$$  |z_j \rangle = N  e^{z_j a^\dag_j } |0\rangle  \ \ \ ; \ \ \ H_j =  \alpha_j (a_j^\dag - \bar{z}_j) + \bar{\alpha}_j(a_j - z_j)       $$
will yield identical results to above.  This is just a shift of the original subsystem Hamiltonian.  If we were to use the original Hamiltonian along with the coherent state reference state, this will thus affect the $a_n$ coefficients.  However, this will be equal to $a_{j,n} = \alpha_j \bar{z}_j + \bar{\alpha}_j z_j $ for all $n$ and the complexity depends only on the difference of the $a_n$ coefficients.  We thus conclude that the excess complexity for any state on the Krylov subspace sourced by a Glauber coherent state is $0$.  \\ \\
The example above corresponds to Lanczos coefficients 
\begin{eqnarray}
a_{j,n} - a_{j,0} & = & 0    \nonumber \\
b_{j,n} & = &   \alpha_j \sqrt{n}. \nonumber
\end{eqnarray}
In our expressions at early times, it may be verified that these choices of the Lanczos coefficients give rise to a vanishing excess complexity.  This may be understood in another way.  The complexity for such a Hamiltonian is quadratic in time (see e.g. \cite{Caputa:2021sib,Balasubramanian:2025xkj}).  Since the excess complexity is $O(t^4)$, it is obvious that this example should give rise to a vanishing spread complexity.  \\ \\
A closely related example that gives rise to a vanishing excess complexity is the Hamiltonian
$$  H = \sum_{j} \alpha_j^\dag a_j^\dag +  \sum_{j} \alpha_j^\dag a_j^\dag + \gamma \sum_{j} \gamma_j a_j^\dag a_j    $$
with the lowest weight reference state.  This corresponds to 
\begin{eqnarray}
a_{n,j} - a_{0,j} & = & \gamma n    \nonumber \\
b_{n,j} & = &   \alpha_j \sqrt{n} \nonumber
\end{eqnarray}
which also gives rise to a vanishing excess complexity.

\section{Pauli spin matrix representation of $su(2)$ elements}
\label{Pauli}
We set out to simplify $\phi_{0,n} = \langle j,-j | e^{itH_n} | j,-j \rangle$ by imposing effectively a basis expansion of $SU(2)$ group elements 
$e^{itH_{n}}=e^{\alpha J_+^{(k_n)}}e^{\gamma J_0^{(k_n)}}e^{\alpha' J_-^{(k_n)}}.$ It is important to note that the parameters attached to each $su(2)$ algebra element differ depending on $n$ as seen in Section~\ref{Kitaev}. One can see that the $e^{\alpha J_+^{(k_n)}}$ and $e^{\alpha' J_-^{(k_n)}}$ factors were chosen so as to act effectively as identity factors within the inner product. All that remains is then to calculate $\gamma_n.$ To do this, make the explicit choice of representation
\begin{eqnarray*}
J_-^{(k_n)} \;\mapsto\;
\begin{pmatrix}
0 & 0\\
1 & 0
\end{pmatrix},
\quad
J_+^{(k_n)} \;\mapsto\;
\begin{pmatrix}
0 & 1\\
0 & 0
\end{pmatrix},
\quad
J_0^{(k_n)} \;\mapsto\;
\begin{pmatrix}
\frac{1}{2} & 0\\
0 & -\frac{1}{2}
\end{pmatrix}.
\end{eqnarray*}
After substituting this into the basis expansion and writing each of the exponential factors on the right-hand side as a Taylor series, some simplification then brings us to 
\begin{eqnarray*}
e^{\alpha J_+^{(k_n)}}e^{\gamma J_0^{(k_n)}}e^{\alpha' J_-^{(k_n)}}=
\begin{pmatrix}
e^{\gamma/2} + \alpha\alpha' e^{-\gamma/2} & \alpha e^{-\gamma/2}\\
\alpha' e^{-\gamma/2} & e^{-\gamma/2}
\end{pmatrix}.
\end{eqnarray*}
Substitution also yields 
$$itH_{k_n}=\begin{pmatrix}
 -it(\mu + J\cos (k_n)) & -t\Delta\sin(k_n) \\
t\Delta\sin(k_n) & it(\mu + J\cos (k_n))
\end{pmatrix}.$$ Verifying that this matrix is traceless, and implementing Equation (2.22) from \cite{Hall2003} gives that $$e^{-itH_{k_n}}
=
\cos\!\left(\sqrt{\det(-itH_{k_n})}\right)\,I
+
\frac{\sin\!\left(\sqrt{\det(-itH_{k_n})}\right)}{\sqrt{\det(-itH_{k_n})}}\,
(-itH_{k_n}).$$ Expanding the determinant then gives 
\begin{eqnarray*}e^{itH_{k_n}}
&=
C_{k_n}(t)I
+it
S_{k_n}(t)
H_{k_n},\end{eqnarray*}with $C_{k_n}$ and $S_{k_n}$ defined in Section~\ref{Kitaev}. Then, equating the above to its (matrix form) basis-expansion equivalent gives
{\footnotesize
\begin{align*}
\begin{pmatrix}
C -iS(\mu+J\cos(k_n)) & S\Delta\sin(k_n)\\
-S\Delta\sin(k_n) & C+iS(\mu+J\cos(k_n))
\end{pmatrix}
=
\begin{pmatrix}
e^{\gamma/2} + \alpha\alpha' e^{-\gamma/2} & \alpha e^{-\gamma/2}\\
\alpha' e^{-\gamma/2} & e^{-\gamma/2}
\end{pmatrix}.
\end{align*}
}
Isolating the (2,2) element of this equation and solving for $\gamma$ (subscripted below to indicate $k_n$-dependence) then results in the form $$\gamma_{k_n}(t)=-\ln\left\{[C_{k_n}(t)+i[\mu+J\cos(k_n)]S_{k_n}(t)]^{2}\right\},$$ which naturally leads to 
\begin{eqnarray}
\phi_{0,n}(t) & = & \exp(s\ln\left\{[C_{k_n}(t)+i[\mu+J\cos(k_n)]S_{k_n}(t)]^2\right\})     \nonumber \\
&=&\{C_{k_n}(t)+i[\mu+J\cos(k_n)]S_{k_n}(t)\}^{2s}.    \nonumber
\end{eqnarray}

\bibliographystyle{elsarticle-num} 
\bibliography{biblio}






\end{document}